\documentclass[runningheads]{llncs}
\usepackage{amsmath,amsfonts,bm}

\def\eqref#1{equation~\ref{#1}}

\def\1{\bm{1}}

\DeclareMathAlphabet{\mathsfit}{\encodingdefault}{\sfdefault}{m}{sl}
\SetMathAlphabet{\mathsfit}{bold}{\encodingdefault}{\sfdefault}{bx}{n}

\newcommand{\R}{\mathbb{R}}

\newcommand{\sigmoid}{\sigma}

\usepackage[ruled,vlined]{algorithm2e}
\usepackage{import}
\usepackage{paralist}
\usepackage{subcaption}
\usepackage{booktabs}
\usepackage{hyperref}
\usepackage{caption}
\usepackage{multirow}
\usepackage{makecell}
\usepackage{threeparttable}
\usepackage{graphicx} 
\usepackage{amsmath,amsfonts,amssymb}
\begin{document}
%
\title{Intention Adaptive Graph Neural Network for Category-aware Session-based Recommendation}
%

\newcommand{\repeatthanks}{\textsuperscript{\thefootnote}}

\author{Chuan Cui\inst{1}\thanks{Both authors contributed equally to this research.} \and
Qi Shen\inst{1}\repeatthanks \and
Shixuan Zhu\inst{1} \and
Yitong Pang\inst{1} \and
Yiming Zhang\inst{1} \and
Hanning Gao\inst{1} \and
Zhihua Wei\inst{1}\thanks{Corresponding author.}
}
\institute{Tongji University, Shanghai, China
\email{\{2033065,2130777,2130768,1930796,2030796,2030795,zhihua\_wei\}@tongji.edu.cn}}
%
\authorrunning{C. Cui, Q. Shen et al.}
\titlerunning{IAGNN for Category-aware Session-based Recommendation}
\maketitle              
\begin{abstract}

Session-based recommendation (SBR) is proposed to recommend items within short sessions given that user profiles are invisible in various scenarios nowadays, such as e-commerce and short video recommendation. 
There is a common scenario that user specifies a target category of items as a global filter, however previous SBR settings mainly consider the item sequence and overlook the rich target category information.
Therefore, we define a new task called Category-aware Session-Based Recommendation (CSBR), focusing on the above scenario, in which the user-specified category can be efficiently utilized by the recommendation system.
To address the challenges of the proposed task, we develop a novel method called Intention Adaptive Graph Neural Network (IAGNN), which takes advantage of relationship between items and their categories to achieve an accurate recommendation result. 
Specifically, we construct a category-aware graph with both item and category nodes to represent the complex transition information in the session.
An intention-adaptive graph neural network on the category-aware graph is utilized to capture user intention by transferring the historical interaction information to the user-specified category domain.
Extensive experiments on three real-world datasets are conducted to show our IAGNN outperforms the state-of-the-art baselines in the new task.\footnote{Source code can be found here: \url{https://github.com/strawhatboy/IAGNN}}

\end{abstract}

\section{Introduction}
\textit{Recommender systems} (RS) have become indispensable nowadays in scenarios such as online shopping or social media, to provide users with accurate information in an effective way. Thanks to the development in deep neural networks, more and more powerful RS methods have been proposed.
Most of them assume the user profile and historical interactions are well recorded.
Nevertheless, many services allow user interaction without user identification, therefore \textit{session-based recommendation} (SBR) was proposed specially. Most of them are based on \textit{recurrent neural networks} (RNNs) \cite{gru4rec,narm} or \textit{graph neural networks} (GNNs) \cite{srgnn,stargnn}. 

\begin{figure*}
    \centering
    \includegraphics[width=0.74\textwidth]{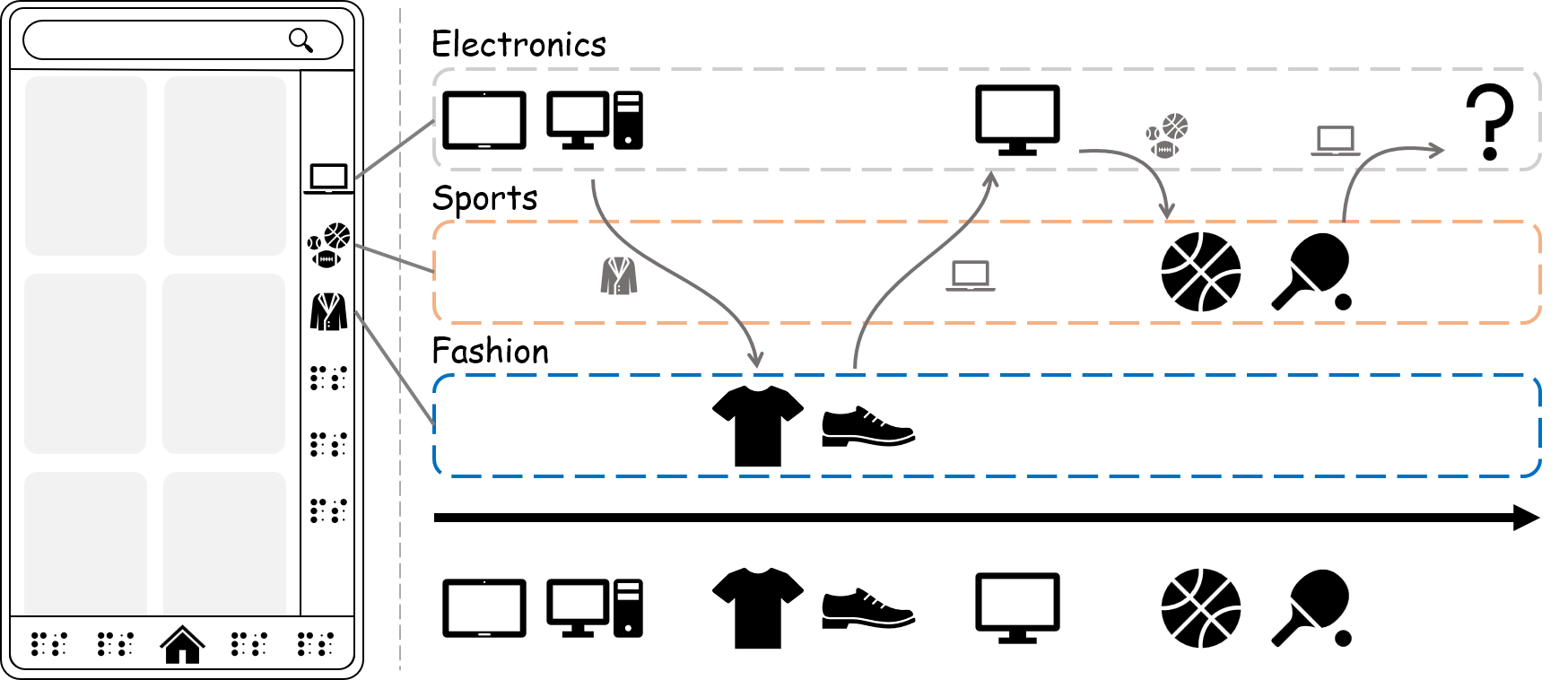}
    \caption{A toy example of Category-aware Session-based Recommendation. }
    \label{fig:toy_example}
\end{figure*}
Previous SBR settings usually focus on the interacted item sequence without consideration of users' other types of behavior, e.g., clicking the specific category button to filter out items.
However, this kind of behavior truely exists in a real-world SBR scenario.
For instance, \autoref{fig:toy_example} illustrates a regular page that allows specifying a category to filter items, with the selection buttons on the navigation bar. During an online shopping trip, after several browsing among different categories, user may wonder if there is any interesting item in the category ``Electronics''.  Therefore he taps the ``Electronics'' button and interacts with items in this target category.
This implies the following possible user demands:
(i) The recommendation direction deviates from the user's intention, therefore user wants to adjust the recommendation results via additional feedback in time, e.g., specifying the target category in accord with his intention.
(ii) Sometimes the recommendation results could not keep up with user's dynamic and abruptly shifting intention, therefore he may want to filter the items with target category to locate the right item efficiently.
(iii) During the interaction, user may tend to concentrate on items in one category than the others, and an autonomous category selection by the user becomes necessary.
In summary, these scenarios require the system to model the dynamic and rich interaction sequence (including interactions with item and category), to make more accurate and user-controllable recommendations based on user-specified target category, rather than general recommendations under all categories.
When facing above scenarios, previous SBR settings cannot perceive the specific category which user intends to view next, and the system fails in reacting to the shifted interest promptly, leading to the failure of meeting user expectations.

Due to above limitations, we propose a new task for SBR: In an ongoing session, with user-specified target category and each item's corresponding category, the system recommends item specifically in the target category. 
The category selection can be commonly seen in the navigation bar of online applications, e.g., the page of an e-commerce platform in \autoref{fig:toy_example}.
This task, namely Category-aware Session-Based Recommendation (CSBR), extends the interaction mode of users, \textit{w.r.t.} specifying target category of next items, to accomplish a user-controllable SBR.


As for the SBR with target category, the key challenge is how to efficiently leverage the category information in addition to the original interaction sequence. 
A straightforward adaption method is to recommend the next item based on the interaction history via existing SBR models and then filter the results by the target category. 
However, this approach overlooks something which are the challenges for CSBR listed as follows:
(i) How to inject the auxiliary category information into session representation dynamically. 
Compared with previous SBR methods, the additional categories information needs to be further considered, including category-level user interaction transition and item-category relations.
(ii) How to transfer historical interaction information to the user-specified category domain efficiently. Intuitively, the interaction in one category might be helpful to improve recommendation effectiveness in other categories, for the idea that user behavior in different categories might reveal a particular user characteristics or interest.
For example, as shown in \autoref{fig:toy_example}, a user might firstly interact with several items in the order of categories ``Electornics'', ``Fashion'', ``Electronics'', ``Sports'' and then he clicks the ``Electronics'' category button to get the recommendation results only in ``Electronics'' category. During this process, some fashion item like ``T-shirt with Super Mario print'' may be helpful to predict the next item ``Nintendo console'' in the ``Electronics'' category because they are all related to the video game ``Mario'' series.
However, adopting user interaction information from other categories will incorporate helpful but also noisy information at the same time. For instance, a ``Swimming google'' in ``Sports'' category might be irrelevant to a game console.
Therefore, it is crucial to distill intention-adaptive information for matching the current interest of user.

To effectively address the aforementioned challenges, we propose a novel model named Intention Adaptive Graph Neural Network (IAGNN) for CSBR. 
In detail, we start by converting the complex interaction session into an Category-aware Graph, which is composed of item-level transitions and category-level transitions, and item-category relations. 
Besides, to transfer historical interaction information to the user-specified category domain, we instantiate the message path between target category node and item nodes with explicit links in the graph. 
Based on the lossless graph, we employ a position-aware graph neural network to learn the item-level and category-level representations.
By stacking multiple layers, the complicated historical interaction information is aggregated to refine the user intention representation in target category iteratively.
Finally, the user's intention is characterized by the target category and last item representations for CSBR.

Our main contributions of this work are summarized below:
\begin{itemize}
    \item We consider a flexible session-based recommendation scenario when user's autonomous navigation in possible item category is perceived, which is introduced as the task \textbf{CSBR}, concentrating on leveraging the signals of user-specified target category and item interaction sequence, for more precise recommendation results.
    \item To address the challenges of CSBR, we propose a novel model named Intention Adaptive Graph Neural Network (\textbf{IAGNN}). Firstly, we construct a category-aware graph not only explicitly models the historical transitions and item-category relations, but also covers the connections of user-specified target category and historical interacted items. Moreover, we conduct a position-aware intention-adaptive graph neural network on the category-aware graph to capture user intention by transferring the historical interaction information to the user-specified category domain.
    \item Extensive experiments on three datasets demonstrate that our model is superior compared with state-of-the-art models for CSBR.
\end{itemize}

\section{Related Works}
In this section, we review some related works of session-based recommendation and category-based recommendation.

\subsection{Session-based Recommendation}
Various neural network based approaches have been proposed for session-based recommendation(SBR) with the development of deep learning recently.

\textbf{RNN-based SBR.}
Recurrent neural networks (RNNs) are powerful sequence models which have been widely adopted for SBR tasks \cite{gru4rec,narm,liu2018stamp,2020s3,ren2019repeatnet,song2019islf}. 
GRU4Rec \cite{gru4rec} by Hidasi et al. was the first RNN-based method to capture information in user-item interaction sequences by simply utilizing several GRU layers. 
NARM \cite{narm} takes advantage of attention mechanism beyond GRU4Rec, by referring the last interacted item, and captures user's preference representations both from global and local perspective of current session. 
However, these sequential methods are unable to capture the items transition relationship efficiently.

\textbf{GNN-based SBR.}
Recently, graph neural networks (GNNs) has been proved to be competent to extract complex relationships between objects, so there emerges quite a few GNN-based methods for SBR \cite{srgnn,xu2019graph,fgnn,gce-gnn,pang2021session,shen2021mb}.
SR-GNN \cite{srgnn} is the first work which leverages gated GNN (GGNN) on a directed graph constructed from the interaction sequence to learn item embeddings.
Nevertheless, SR-GNN only propagates messages between adjacent items, which would fail to take long-distance item relations into consideration.
LESSR \cite{lessr} proposes a better architecture, including EOPA layer and SGAT layer to solve the lossy session encoding problem and propagate information along shortcut connections, which leads to lossless information presentation during graph construction and better performance for SBR.
Pan et al. proposed StarGNN \cite{stargnn} with a star graph neural network to model the complex transition relationship between items with an additional star node connected to every item in an on going session, and applies a highway network to handle the over-fitting issue in GNNs.

\subsection{Category information in Recommendation}
As a  significant auxiliary information for items, category information has been explored in other recommendation areas. 

\textbf{Cross-domain recommendation.} Cross-domain recommendation (CDR) \cite{conet,kang2019semi} can be considered as general multi-category recommendation, which utilizes data from multiple domains to deal with issues like cold start \cite{abel2013cross} and data sparsity \cite{pan2010transfer} in target domain (category). 
Recently, Ma et al. proposed $\pi$-net \cite{ma2019pi} to generate recommendation scores for every item in two domains by integrating information from both.
DA-GCN \cite{guo2021gcn} constructs a cross-domain sequence graph which explicitly links account-sharing users and items from two domains to learn expressive representations for recommendation.

\textbf{Category-aware recommendation.}
Category-aware recommendation utilizes category information to enhance the item representation for better user preference modeling.
There are several traditional category-aware recommendation methods. HRPCA \cite{albadvi2009hybrid} proposes a hybrid recommendation method to handle the customer preferences varieties in different product categories. 
Choi et al. \cite{choi2010content} design a recommendation algorithm  based on  the category correlations to a user with certain preferences.
Recently, LBPR \cite{he2017category} conducts list-wise bayesian ranking for next category and category-based location recommendation. CoCoRec \cite{cai2021category} utilizes self-attention to model item transition patterns in category-specific action sub-sequences, and  recommends items for user by collaborating neighbors' in-category preferences.

As outlined above, the rich category information has not been comprehensively explored by previous SBR methods.
Also for previous category-related methods, they rarely focused on recommendation for a session with anonymous users.
What is more, none of these task settings explicitly consider the user-specified  category information, and there are huge gaps in this scenario.

\section{Preliminary}\label{sec:define}

Given the entire item set $\mathcal{V}$ and category set $\mathcal{C}$, we first define a category-augmented item session as $s=\{(v_1,c_{v_1}),(v_2,c_{v_2}),...,(v_n,c_{v_n})\}$, in which $(v_i,c_{v_i})$ represents the user interacted item $v_i \in \mathcal{V}$ in category $c_{v_i} \in \mathcal{C}$, and $n$ is the session length. Note that $v_i$ and $c_{v_i}$ can be repetitive in the sequence, since there can be repeated items in the session.
Given session $s$, the user-specified target category $c_t$ and the corresponding item set $\mathcal{V}_{c_{t}}=\{v_i|c_{v_i}=c_t,v_i \in \mathcal{V}\}$, the goal of CSBR is to predict a probability score for any item $v \in \mathcal{V}_{c_{t}}$ such that an item with higher score is more likely to be interacted next.

\begin{figure*}
    \centering
    \includegraphics[width=0.95\textwidth]{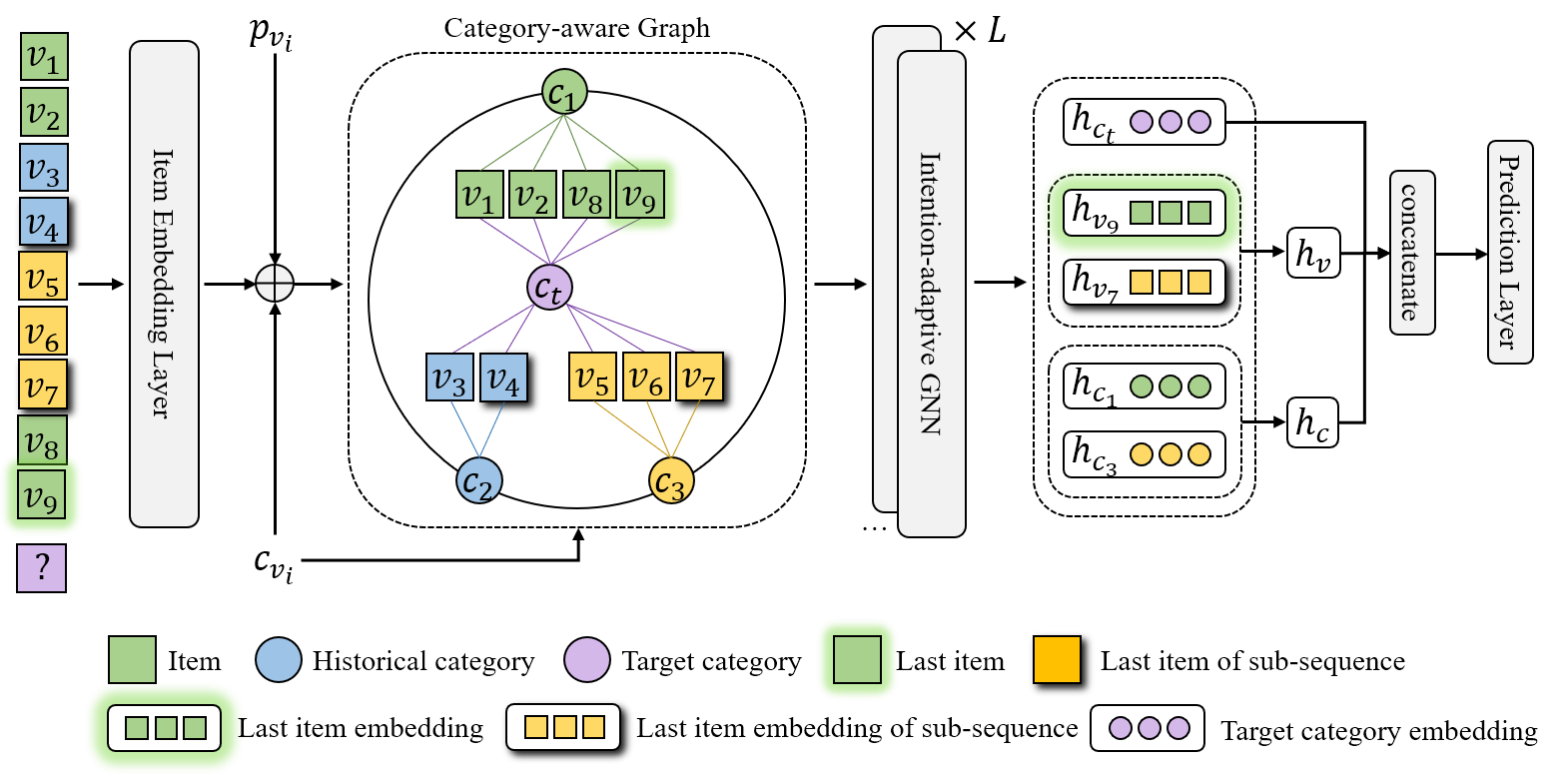}
    \caption{The overview of IAGNN. }
    \label{fig:framework}
\end{figure*}

\section{Methodology}\label{sec:model}
\subsection{Overall Architecture}
\autoref{fig:framework} illustrates the overall architecture of IAGNN under the context of user's target category specified.
First, the Embedding Layer will initialize id embeddings for all items and categories. 
Second, we construct the Category-aware Graph to explicitly keep the transitions of in-category items and different categories, along with the relation between items and their corresponding categories.
Third, a Heterogeneous Graph Attention Neural Network is introduced to propagate embeddings through attentive graph convolution. Finally, by leveraging a Embedding Fusion Layer and a Prediction Layer, items in user-specified target category will be recommended.

\subsection{Category-aware Graph Construction}
According to previous SBR methods, category information is not explicitly integrated in the recommendation process. In our proposal, we take advantage of graph neural network to model the items and categories transition. The graph $\mathcal{G}_s = (\mathcal{V}_s,\mathcal{E}_s)$ is a directed heterogeneous graph, in which $\mathcal{V}_s = (\{v_1, v_2, v_3,...,v_n\}, \{c_1, c_2, c_3,...,c_h\}, c_t )$ indicates $n$ item nodes, $h$ category nodes and the user-specified target node $c_t$ in the graph. $\mathcal{E}_s$ is comprised of the edges representing item transitions, category transitions, item-category connections and target category connections. Therefore, the semantics of item transition and category transition can be represented by these nodes and their relations.
Generally, there are four sub-graphs with independent edge relations, and we will detail them as follows.

\textbf{Item Transitions.}
At first, we split the original interaction sequence into several category-specific item sub-sequences. For instance, items in category $c$ will be connected with their original interaction order in the session, and the same for other categories. Therefore, the in-category item-to-item transition patterns can be modeled in the sub-sequence.


In detail, for each item node $v_i$ in the category $c_{v_i}$, we create a link starting from $v_i$ to its next item $v_j$, representing the item transition in the category sub-sequence. 
And we also create self-loop for each item node to include the node itself when aggregating.


\textbf{Category Transitions.}
Considering each item node $v_i$ in the original session sequence, we can simultaneously initialize a corresponding category node $c_{v_i}$ for it. And a category transition graph can be constructed according to the order of the original session sequence order. For instance, in \autoref{fig:framework}, $c_1$, $c_2$, $c_3$ indicate the categories in colors `Green', `Blue' and `Yellow'. The session sequence consists of items in category $[c_1,c_2,c_3,c_1]$ respectively, then a sub-graph composed of connections between categories can be constructed by the directed link $c_1\rightarrow c_2\rightarrow c_3\rightarrow c_1$. Similar to the item transition sub-graph, a self-loop is added to every category node $c_i$. 

\textbf{Item-Category Connections.} 
For representing the inherent item-category relations, we build bi-directed connections between item node $v_{i}$ and its corresponding category node $c_{v_i}$.
By this connection, information from non-adjacent items nodes can be propagated in a two-hop way through the corresponding category node as an intermediate node. 
Besides, the category transition information which implies the user interest changes across different categories can be propagated back to all the related item nodes. 
Finally, this approach can enhance the representation of both items and categories.

\textbf{Target Category connections.}
Moreover, the user-specified target category is a given information in our task, to better leverage this information, we add a particular node $c_t$ to represent the target category, which is connected to all the item nodes (In \autoref{fig:framework}, $c_t$ represents the same category with $c_3$).
Therefore, it can not only solve the long-range information propagation issue between non-adjacent items, but also build the transfer from all the historical information to the target category domain as the the process of message passing from items to target category node. 
Meanwhile, as a pseudo node for the next item, the representation of $c_t$ can be an intermediate variable to bridge the gap of all the interacted items and the actual next item in the session. 

\subsection{Intention-adaptive Graph Neural Network}
Next, we present how intention-adaptive graph neural networks propagate messages between different nodes.

\textbf{Embeddings.}
For each item $v_i$, it is projected by the item embedding layer $E_i \in \R^{|V| \times d_{v}}$ into dense embedding $e_{v_i}$. In which $d_{v}$ denotes the item embedding size. Meanwhile the category embedding layer $E_c \in \R^{|C| \times d_{c}}$ transform category unique identifications to category embedding $e_{c_i}$. Note that since $c_t$ represents the target category, it can also be initialized in the same way as a regular category node.

Also for a session with no more than $n$ items, we take advantage of the reversed position embeddings $\mathbf{P}_{-i} \in \R^{n \times d_{P}}$ by following GCE-GNN\cite{gce-gnn}, which shows superiority performance because the session length is versatile. Moreover, we conduct a positive-going position embedding in our experiments. Also the category embeddings $\mathbf{e}_{c_{v_i}}$ are added for each item $v_i$ to enhance the item representation by concatenation and projection:
\begin{equation}
\begin{aligned}
    \mathbf{e}_{v_i} = \mathbf{W}^{'} \left[ \mathbf{e}_{v_i} \parallel \mathbf{e}_{c_{v_i}} \parallel \mathbf{p}_{v_i} \right]
\end{aligned} 
\end{equation}
where $\parallel$ is the concatenation operation, $\mathbf{W}^{'} \in \R^{d_v \times (d_v + d_c + d_P)}$ is the projection matrix to keep the size of $e_{v_i}$, $\mathbf{p}_{v_i} \in \mathbf{P}_{-i}$ is the reversed position embedding for $v_i$.

\textbf{Message propagation.}
Our method uses the same message propagation mechanism for the different edge types in the graph.
We use item-item message propagation as an example here. As mentioned above, the item node $v_i$ is linked to its previous item in the in-category sequence along with the self-loop connection.
Then the item message that node $v_i$ received can be represented as:

\begin{equation} \label{eq:msg1}
\begin{aligned}
    \mathbf{h}_{v_i \leftarrow v}^{(l+1)} =\sigmoid ( \sum_{j \in \mathcal{N}_{v_i}(v)} \alpha_{ij}^{(l)}    \mathbf{W}_v \mathbf{h}_{v_j}^{(l)} )
\end{aligned} 
\end{equation}
\begin{equation} \label{eq:msg2}
\begin{aligned}
    \alpha_{ij} = \frac {\mathrm{exp}(\mathrm{LeakyReLU} ( \mathbf{a}_{l_v} \mathbf{W}_v \mathbf{h}_i + \mathbf{a}_{r_v} \mathbf{W}_v \mathbf{h}_j))} { \sum_{k \in \mathcal{N}_{v_i}(v)} \mathrm{exp} (\mathrm{LeakyReLU} ( \mathbf{a}_{l_v} \mathbf{W}_v \mathbf{h}_i + \mathbf{a}_{r_v} \mathbf{W}_v \mathbf{h}_k)) }
\end{aligned}
\end{equation}
where $\mathbf{h}_{v_i \leftarrow v}^{(l+1)} \in \R^{d_{v}}$ is the item-source message representation of node $v_i$ at layer $(l+1)$ and $\mathcal{N}_{v_i}(v)$ stands for the set of neighbor item nodes. 
$\alpha_{ij}$ is a scalar representing the attention coefficient. $\mathbf{W}_v \in \R^{d_v \times d_v}$ and $\mathbf{a}_{l_v}, \mathbf{a}_{r_v} \in \R^{1 \times d_v}$ are the shared linear transformation parameters for the source and target node.

\textbf{Graph nodes aggregation.}
Similar to the item-item message aggregation (\autoref{eq:msg1} and \ref{eq:msg2}), we can get other types of messages from the adjacent nodes. Therefore, the final representation for an item node at layer $(l+1)$ can be represented as:
\begin{equation} \label{eq:agg_hv}
\begin{aligned}
    \mathbf{h}_{v_i}^{(l+1)} = \mathbf{h}_{v_i \leftarrow v}^{(l+1)} + \mathbf{h}_{v_i \leftarrow c}^{(l+1)} + \mathbf{h}_{v_i \leftarrow c_t}^{(l+1)}
\end{aligned} 
\end{equation}
where $\mathbf{h}_{v_i \leftarrow v}$ denotes the messages passed from other item nodes, $\mathbf{h}_{v_i \leftarrow c}$ represents the messages from its corresponding category node, and $\mathbf{h}_{v_i \leftarrow c_t}$ stands for the messages from target category.

As for the category nodes, its representation at layer $(l+1)$ is:
\begin{equation} \label{eq:agg_hci}
\begin{aligned}
    \mathbf{h}_{c_i}^{(l+1)} = \mathbf{h}_{c_i \leftarrow v}^{(l+1)} + \mathbf{h}_{c_i \leftarrow c}^{(l+1)}
\end{aligned} 
\end{equation}
In which $\mathbf{h}_{c_i \leftarrow v}^{(l+1)}$ is the messages propagated from adjacent item nodes, and $\mathbf{h}_{c_i \leftarrow c}^{(l+1)}$ represents the messages from adjacent category nodes.

The same for the target category node $c_t$, its $(l+1)$ layer representation is:
\begin{equation} \label{eq:agg_hct}
\begin{aligned}
    \mathbf{h}_{c_t}^{(l+1)} = \mathbf{h}_{c_t \leftarrow v}^{(l+1)}
\end{aligned} 
\end{equation}
Since the target category node connects to all the item nodes, only $\mathbf{h}_{c_t \leftarrow v}^{(l+1)}$, the messages aggregated from item nodes are used for its representation.

As we employs an $L$-layer IAGNN, the embedding of each node before and after the IAGNN can be represented as $\mathbf{h}^{(0)}$ and $\mathbf{h}^{(L)}$ respectively. In order to fuse the semantics before and after the network and avoid possible overfitting, we utilize the \textit{gate} mechanism to get the final embeddings for each kind of node:
\begin{equation} \label{eq:gate1}
\begin{aligned}
    \mathbf{h}^L = \mathbf{g} \odot \mathbf{h}^{(0)} + (1-\mathbf{g}) \odot \mathbf{h}^{(L)}
\end{aligned} 
\end{equation}
\begin{equation} \label{eq:gate2}
\begin{aligned}
    \mathbf{g} = \sigmoid ( \mathbf{W} [ \mathbf{h}^{(0)} \parallel \mathbf{h}^{(L)} ] )
\end{aligned} 
\end{equation}
where the $\mathbf{g}$ is the gate factor to control information contributed from $\mathbf{h}^{(0)}$ and $\mathbf{h}^{(L)}$. $\mathbf{W} \in \R^{(d_h \times 2) \times d_h }$ is the transform matrix to get the gate. $\sigmoid$ denotes the \textit{Sigmoid} activation function. Here we use a function \textit{Gated} to simplify \autoref{eq:gate1} and \ref{eq:gate2}:
\begin{equation} 
    \begin{aligned}
        \mathbf{h}^L = \mathrm{Gated}\left(\mathbf{h}^{(0)} , \mathbf{h}^{(L)}\right)
    \end{aligned} 
    \end{equation}

\subsection{Embeddings fusion and Prediction}
In a SBR scenario, the last item user interacted contributes a lot to the next item user will interact with. In our case, there are two last items, one is the last item $v_l$ of the original interaction sequence, the other is the last item $v_{l_c}$ of the in-category sub-sequence. Therefore, after going through the IAGNN, we fuse these two item representations to a session item representation:
\begin{equation} \label{eq:fused_hv}
\begin{aligned}
    \mathbf{h}_{v} = \mathrm{Gated}\left(\mathbf{h}_{v_l}^L, \mathbf{h}_{v_{l_c}}^L\right)
\end{aligned} 
\end{equation}

Accordingly, we fuse the category node embeddings $c_l$ and $c_{l_c}$ which is related to the sequence last item $v_l$ and last item $v_{l_c}$ of in-category sub-sequence as the session category representation:
\begin{equation} \label{eq:fused_hc}
\begin{aligned}
    \mathbf{h}_{c} = \mathrm{Gated}\left( \mathbf{h}_{c_l}^L, \mathbf{h}_{c_{l_c}}^L\right)
\end{aligned} 
\end{equation}

After obtaining the item and category session representation $h_v$ and $h_c$, we can also get the embedding of target category node:
\begin{equation} \label{eq:fused_hct}
\begin{aligned}
    \mathbf{h}_{c_t} = \mathbf{h}_{c_t}^L
\end{aligned} 
\end{equation}
Then, we combine them into one embedding for the session:
\begin{equation} \label{eq:session_emb}
\begin{aligned}
    \mathbf{h}_s = \mathbf{W}_{s}\left[\mathbf{h}_v \parallel \mathbf{h}_c \parallel \mathbf{h}_{c_t}\right]
\end{aligned} 
\end{equation}
where $\mathbf{W}_{s} \in \R^{d_v \times (d_v + d_c \times 2)}$ is used to project the concatenation result to an embedding of size $d_v$. Note that we did not introduce additional attention mechanism for all the items as a readout step, because regarding the target category node, the attention mechanism is equivalently done after message propagation through multi-layer GNN.

By multiplying the session representation with all item embeddings, we can get the prediction score as follows:
\begin{equation}
\begin{aligned}
    \hat{y_i} = \mathbf{h}_s^{\intercal} \tilde{\mathbf{e}_{v_i}}
\end{aligned} 
\end{equation}
where $\tilde{e_{v_i}}$ is the projected candidate item embeddings through item embedding layer $E_i \in \R^{|V| \times d_{v}}$.

To train our network, a cross-entropy loss function is employed to optimize the model parameters:
\begin{equation}
\begin{aligned}
    L(\hat{\mathbf{y}}) = - \sum_{i=1}^{|V|}y_i \mathrm{log} (\hat{y_i}) + (1 - y_i) \mathrm{log} (\hat{y_i})
\end{aligned} 
\end{equation}
where $y_i$ denotes the ground-truth item one-hot encoding vector.


\section{Experiments}
We conduct extensive experiments to evaluate our method in comparison with other state-of-the-art methods. The goal in this section is to answer the following research questions:
\begin{itemize}
    \item \textbf{RQ1:} What is the performance difference between SOTA baselines and ours? 
    \item \textbf{RQ2:} Can the category information of the sequence items contribute to the recommendation performance?
    \item \textbf{RQ3:} Will the graph construction benefit the model performance?
    \item \textbf{RQ4:} How does our model perform on the task CSBR with different hyper-parameters setup?
\end{itemize} 

\newcommand{\tabincell}[2]{\begin{tabular}{@{}#1@{}}#2\end{tabular}}  
\begin{table}[htbp]
    \caption{Statistics of datasets used in experiments.}
    \label{tab:dataset}
    \centering
    \begin{tabular}{lrrr}
    \toprule
    Statistic& $\text{Diginetica}^{\star}$ & $\text{Yoochoose}^{\star}$ & $\text{Jdata}^{\star}$ \\
    \midrule
    No. of items  & 33,596 & 13,317  & 79,356 \\
    No. of sessions  & 482,743 & 400,613 & 959,824 \\
    Avg. of session length & 6.48 & 10.27 &  6.90\\
    No. of categories &982 &12 &79 \\
    Avg. of categories per session &2.28 &2.07 &2.35 \\
    \bottomrule
    \end{tabular}
\end{table}

\subsection{Experimental Setup}
\textbf{Dataset.} 
We performed the evaluation on three public datasets: $\text{Diginetica}^{\star}$, $\text{Yoochoose}^{\star}$ and $\text{Jdata}^{\star}$, which are widely used in the session-based recommendation research \cite{lessr,gce-gnn,MKMSR}. 
These datasets contain additional category information which can support our work for CSBR.

\begin{itemize}
    \item $\text{Diginetica}^{\star}$\footnote{\url{https://competitions.codalab.org/competitions/11161}} includes user sessions extracted from e-commerce search logs, with desensitized user ids, hashed queries, hashed query terms, hashed product descriptions and meta-data, log-scaled prices, clicks, and purchases.
    

    \item $\text{Yoochoose}^{\star}$\footnote{\url{https://www.kaggle.com/chadgostopp/recsys-challenge-2015}} is the dataset for RecSys Challenge 2015, which contains user clicks and purchases of an online retailer within several months. In our case, we use the fractions 1/4 of Yoochoose data.
    
    \item $\text{Jdata}^{\star}$\footnote{\url{https://jdata.jd.com/html/detail.html?id=8}} is also a dataset of a challenge hosted by JD.com. The data was filtered by 1 hour duration to extract the session data.
\end{itemize}

To better apply the datasets to our task, we pre-process them before training. 
At first, as described in \cite{narm,liu2018stamp,srgnn}, we applied a data augmentation by regarding the $i\mathrm{th}$ item as the label and items before $i$ as the input sequence. 
Moreover, the category of the label item is considered as the user-specified target category.
Then, since there could be short sessions or infrequent items, we removed all sessions of length $\leq2$ and items which have an occurrence less than 5 times in all datasets.
In the end, for the given specific category in one session, we filtered the candidate items by keeping the category of candidate item the same as the target category.
We summarized the preprocessed dataset in \autoref{tab:dataset}.

\textbf{Baseline Models.}
To evaluate the performance of our method, we compare it with several baselines.
Note that the candidate items in the experiments are all filtered with the target category, which means we evaluate the precision and rank performance only on the candidate items in the target category for all models.


\begin{itemize}
    \item \textbf{GRU4Rec}\cite{gru4rec} captures patterns in user-item interaction sequences by simply utilizing several GRU layers.
    \item \textbf{NARM} \cite{narm} combines the global and local representations to generate the session embedding through attention mechanism and RNN model.
    \item \textbf{SR-GNN} \cite{srgnn} transforms user interaction sequences into directed graphs and utilizes the GGNN layer\cite{li2015gated} to learn the embeddings of sessions and items.
    \item \textbf{LESSR} \cite{lessr} uses a lossless GRU and a shortcut graph attention layer to capture long-range dependencies and lossless embeddings.
    \item \textbf{StarGNN} \cite{stargnn} employs a additional star graph neural network to model the complex transition in sessions.
    \item $\textbf{DA-GCN}^{\dagger}$ \cite{guo2021gcn} uses a domain-aware GNN and two novel attention mechanisms to learn the sequence representation. Here we change the domain concept of the original DA-GCN to category in $\text{DA-GCN}^{\dagger}$ by employing in-category sub-sequences, along with the target category integrated.
    
\end{itemize}

Nevertheless, we adopt the target category information to NARM, LESSR, StarGNN by fusing the target category embedding to the query part of their attention mechanism, named $\text{NARM}^{\dagger}$, $\text{LESSR}^{\dagger}$ and $\text{StarGNN}^{\dagger}$. 

\begin{table*}[t]
    \centering
    \caption{Experimental results (\%) of different models in terms of P@\{10, 20\}, and mrr@\{10, 20\} on three datasets. The * means the best results on baseline methods. $Improv.$ means improvement over the state-of-art methods. The bold number indicates the improvements over the strongest baseline are statistically significant ($p \textless 0.01$) with paired t-tests.}
    \label{tab:overall}
    \resizebox{\textwidth}{27mm}{
    \begin{tabular}{p{1.55cm}<{\centering}p{0.89cm}<{\centering}p{0.92cm}<{\centering}p{1.05cm}<{\centering}p{1.0cm}<{\centering}p{0.05cm}p{0.89cm}<{\centering}p{0.92cm}<{\centering}p{1.05cm}<{\centering}p{1.0cm}<{\centering}p{0.05cm}p{0.89cm}<{\centering}p{0.92cm}<{\centering}p{1.05cm}<{\centering}p{1.0cm}<{\centering}}
    \toprule
    \multirow{2}{*}{ \bfseries Models}& \multicolumn{4}{c}{ \bfseries $\text{Diginetica}^{\star}$ }& & \multicolumn{4}{c}{\bfseries $\text{Yoochoose}^{\star}$ }& &\multicolumn{4}{c}{\bfseries $\text{Jdata}^{\star}$} \\
    \cline{2-5}
    
    \cline{7-10}
    
    \cline{12-15}
    &P@10 &P@20 &mrr@10&mrr@20 && P@10& P@20 &mrr@10 &mrr@20  && P@10 &P@20 &mrr@10 &mrr@20\\
    \midrule
     GRU4Rec &27.47 &36.52 &15.02 &15.89  &&10.25  &13.93  &3.92 &4.38  &&31.20  &43.79  &13.91  &14.97 \\
    NARM &28.86 &37.84 &16.89 &17.53  &&11.86  &15.25  &5.98  &6.21  &&33.92 &45.01  &15.52  &16.29   \\
    
    SR-GNN &30.71 &41.86 &16.27 &17.17  &&11.46  &15.39  &4.84  &5.11  &&35.00  &46.30  &16.07  &16.88 \\

    LESSR &30.49 &41.42 &16.52 &17.27  &&11.90  &15.67  &5.85  &6.24  &&35.95  &46.92  &16.95  &17.71  \\
    
    StarGNN &29.86 &40.96 &15.33 &16.09  &&12.46  &17.30  &5.47  &5.80  &&36.91  &48.40  &17.15  &17.94  \\

    \midrule
    $\text{NARM}^{\dagger}$ &28.79 &38.48 &16.95* &17.61  &&11.95  &15.76  &6.28  &6.54  &&33.98  &44.97  &15.61  &16.38  \\
    $\text{LESSR}^{\dagger}$ &30.55 &41.39 &16.77 &17.51  &&12.95  &17.24  &6.20  &6.52  &&36.05  &47.22  &17.01  &17.79  \\
    $\text{StarGNN}^{\dagger}$ &30.67 &41.34 &16.93 &17.66* &&11.73  &16.27  &5.24  &5.55  &&37.21*  &48.75*  &17.53*  &18.33*  \\
    
    $\text{DA-GCN}^{\dagger}$  & 31.49* &42.53* &16.86 &17.62  &&15.53*  &20.15*  &7.09*  &7.41*  &&36.42  &48.30  &16.52  &17.29  \\
    \midrule
    
    IAGNN  &{\bfseries 32.63} &{\bfseries 43.80}& {\bfseries 17.35} &{\bfseries  18.11}& &{\bfseries 17.03}&{\bfseries 21.42} &{\bfseries 8.18}&{\bfseries 8.49} & &{\bfseries  39.86}&{\bfseries 51.79} &{\bfseries  18.89}&{\bfseries 19.72}\\

    {$Improv.$}& 3.62\% & 2.99\% & 2.36\% & 2.55\% & & 9.66\% & 6.30\% & 15.37\% & 14.57\% & & 7.12\% &6.24\% & 7.76\% & 7.58\%    \\
    \bottomrule
    \end{tabular}}
\end{table*}

\textbf{Evaluation Metrics.}
We employ two widely used metrics: \textit{Precision} (P@$k$)  and \textit{Mean reciprocal rank} (mrr@$k$) following \cite{srgnn,lessr}, where $k =\{10, 20\}$. They respectively represent the correct proportion of the top-$k$ result items and the order of recommendation ranking.

\textbf{Implementation Details.}
By leveraging the frameworks PyTorch and Deep Graph Library, we implement our method.
We fix the embedding size of items and categories to $128$. The model parameters are initialized with a Gaussian distribution with $\mu=0$ and $\sigma=0.1$, where $\mu$ and $\sigma$ are the statistical mean and standard deviation.
To help the model converge, We employ the \textit{Adam}\cite{kingma2014adam} optimizer with the mini-batch size of $256$. 
A grid search for the hyper-parameters is the following: learning rate $\eta$ in $\{0.001,0.005,0.01,0.05,0.1\}$, learning rate decay step in $\{2,3,4\}$, number of graph neural network layers $L$ in $\{1,2,3,4,5\}$.
Furthermore, we split the dataset to train, test and validation set by ratio 8:1:1.
Regarding the baseline methods,
we either directly make minimal changes to their original source code for supporting our dataset,
or implement by ourselves according to their papers.

\subsection{Experimental Results (RQ1)}
\autoref{tab:overall} shows the comparison results of IAGNN over other baselines on the preprocessed datasets $\text{Diginetica}^{\star}$, $\text{Yoochoose}^{\star}$ and $\text{Jdata}^{\star}$.

\textbf{Comparison of Different Baselines.} 
The performances of two RNN-based methods, GRU4Rec and NARM, are not so competitive. Nevertheless, NARM performs much better than GRU4Rec because of its usage of attention mechanism to capture the user interest.
And for the GNN-based methods SR-GNN, LESSR and StarGNN, they performs significantly better than the RNN-based methods, which proved that the GNN has promising advantage for SBR.
Moreover, we can discover that LESSR and StarGNN outperform SR-GNN in most cases because they use shortcut graph or the star node to capture global dependencies between distant items, showing the effectiveness of explicitly considering long-range dependencies.

\textbf{Significance of Target Category.} 
Furthermore, for the methods with target category information adopted, $\text{NARM}^{\dagger}$, $\text{LESSR}^{\dagger}$ and $\text{StarGNN}^{\dagger}$ generally outperform their original implementation, which indicates that integrating user-specified target category information can improve the performance significantly for CSBR.
We can notice that $\text{DA-GCN}^{\dagger}$ performs better than the other modified baselines. This means the graph construction method including in-category sub-sequence can benefit our task. Note that for dataset $Diginetica^{\star}$, the overall improvements are not obvious because it has too many categories which can be noisy to the category enhanced models.

\textbf{Model Effectiveness.}
Our model outperforms all the other GNN-based models from \autoref{tab:overall}, which shows our model has a better effectiveness. We can elaborate the advantage of our method in three aspects.
First, we integrate the user-specified target category into the graph as a dedicated node, such that the model can be aware of the target category as the session context.
Second, we inherit the in-category sub-sequence graph construction from the CDR methods, which can model the item transition per category.
Third, we introduce the nodes and transitions for categories, and along with the in-category sub-sequence, we keep the original item sequence which can lead to less information loss.

\begin{table}[htbp]
    \caption{Performance comparison for ablation study.}
    \label{tab:ablation_study}
    \centering
    \setlength{\tabcolsep}{2pt}
    \begin{tabular}{lcccccc}
    \toprule
   
\multirow{2}{*}{\bfseries Model setting} &
  \multicolumn{2}{c}{$\text{Yoochoose}^{\star}$} & &
  \multicolumn{2}{c}{$\text{Jdata}^{\star}$}  &
  \\ \cline{2-3} \cline{5-6}  
    & P@20 &mrr@20 & &P@20 &mrr@20  \\
    \midrule
    {w/o Category nodes} &20.89 &8.12 &&50.84 &19.32 \\
    {w/o Target category information}  &20.91 &8.04	&& 50.87	&19.08	\\
    {w/o Category transition}&21.36 &8.13 &&51.14 &19.47 \\
    \midrule
    {Add Original item transition}&21.22 &8.41 &&51.70  &19.68 \\
    {Add Attention mechanism}&20.76 &7.68 &&50.14  &18.97 \\
    {Positive position information}&21.80 &8.04 &&51.49 &19.62 \\
    
    \midrule
    IAGNN &21.42 &8.49 &&51.79  &19.72 \\
    \bottomrule
    \end{tabular}
\end{table}

\subsection{Ablation Study (RQ2\&3)}\label{sec:ablation}

In this section, we performed some ablation studies to demonstrate the effectiveness of our model designs. 

\textbf{Category information.}
In our proposed method, we created category nodes for in-category sub-sequences during the graph construction. Therefore, we compared our model with the version without sub-sequence category nodes (``w/o Category nodes", by removing $\mathbf{h}_{v_i \leftarrow c}$, $\mathbf{h}_{c}$ in \autoref{eq:agg_hv}, \ref{eq:agg_hci}, \ref{eq:fused_hc} and \ref{eq:session_emb}) in the graph in order to show the effectiveness of introducing these nodes.
On the other hand, to show the importance of introducing user-specified target category information, we used mean pooling of all item representations as the representation of the target category node instead of the target category (``w/o Target category information") embedding to diminish the impact of target category for comparison:
\begin{equation} \label{eq:ect_meanpool}
    \begin{aligned}
        \textbf{e}_{c_t} = \textrm{MeanPool}\left(\textbf{e}_{v_i}\right)
    \end{aligned} 
\end{equation}
Moreover, we compared the case when transitions between each category node were removed (``w/o Category transition", by removing $\mathbf{h}_{c_i \leftarrow c}$ in \autoref{eq:agg_hci}).
As illustrated in \autoref{tab:ablation_study}, the performance dropped on both datasets while detaching category information from the model in different mechanisms. This proved the graph construction details aforementioned in \autoref{sec:model} help improving the model performance.

\begin{figure}[t]
    \centering
    \begin{subfigure}{0.35\linewidth}
        \includegraphics[width=\textwidth]{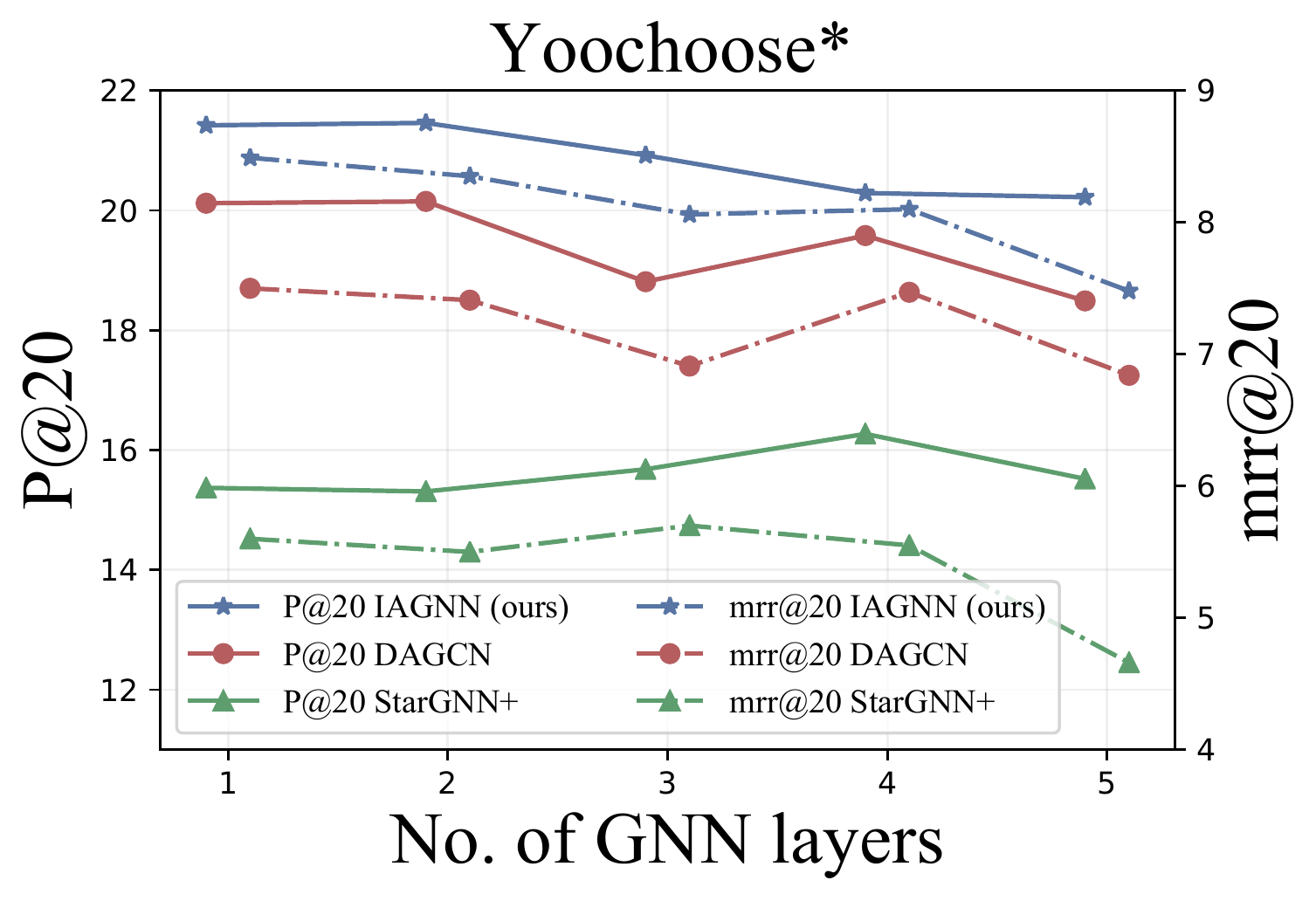}
    \end{subfigure}
    \begin{subfigure}{0.35\linewidth}
        \includegraphics[width=\textwidth]{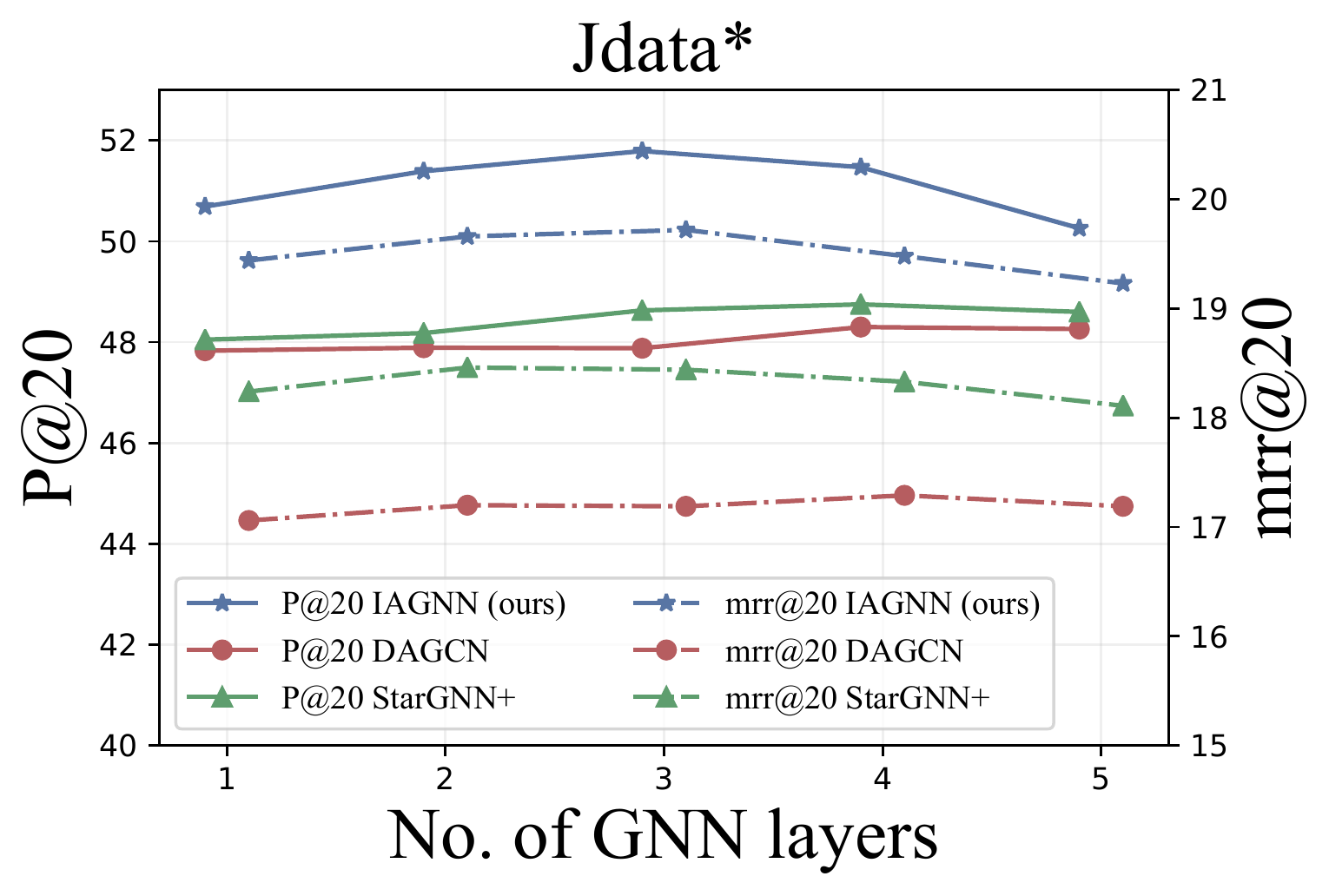}
    \end{subfigure}
    
    \caption{Model comparison w.r.t. different depths of GNN.}
    \label{fig:hyper-depth-compare}
\end{figure}

\textbf{Additional model operations.}
As we know, the items in session has its inherent order, and we added the reversed position embedding for every item node to keep the original item transition information. Thus we tried to add the original item transition as directed links for all the item nodes to check if an additional lossless item order can improve the performance.
Meanwhile, following \cite{narm,srgnn,lessr}, an additional attention mechanism which gets each item contribution was added to our model.
Also we want to see if a positive position embedding is better than the negative. Corresponding to above three conditions, we respectively conduct three alternative models: ``Add Original item transition", ``Add Attention mechanism" and ``Positive position information" for comparison.
Following the result in \autoref{tab:ablation_study}, The performance drops a little by adding original item transition links to the method. The reason is the model already takes advantage of the negative position embedding and category transition which persist the original item transition order, adding these links will be redundant. 
Meanwhile, by introducing attention mechanism, the performance was impaired because the target category information is much more significant than attention on every item from different categories. About the position information, albeit positive position embedding improves for P@20 on dataset $\text{Yoochoose}^{\star}$, overall the performance drops comparing to our method with negative position embedding.

\subsection{Hyper-parameters Study (RQ4) }\label{sec:hyper-res}

\textbf{GNN depths.} We conducted experiments on different number of GNN layers for different GNN-based methods to check the performance impact. As illustrated in \autoref{fig:hyper-depth-compare}, our method keeps beyond the others. Also, we notice that the performance goes up when multi-layer GNN was conducted than that of only one single layer, because more GNN layers are able to capture the high-level and complex semantic information.
Furthermore, the metrics of all methods decline approximately after $4$ GNN layers, which indicates the over-smoothing is a common issue for GNN-based methods. Without exception, the performance of our method drops after about $3$ layers.
Nevertheless, as the number of GNN layers goes up, the performance of almost every method increases at the beginning, because GNN is competent for capturing the node embedding transitions, showing it is suitable for our task which involves item and category transitions.

\section{Conclusion}
This paper proposes a novel category-aware session-based recommendation task, which includes the rich category information both resided in the interaction sequence and the user's intention. Accordingly, a novel model Intention Adaptive Graph Neural Network is introduced to take advantage of these category information, and achieves a more effective performance by explicitly transferring the historical interaction information to the user-specified category domain.
Extensive experiments on three real-world datasets are conducted to show our IAGNN outperforms the state-of-the-art baselines in the new task.

As to future work, we prefer to involve other types of attribute information besides category information which would extend the task to an attribute aware session-based recommendation task. 

\subsubsection{Acknowledgements.}
The work is partially supported by the National Nature Science Foundation of China (No. 61976160, 61976158, 61906137) and Technology research plan project of Ministry of Public and Security (No. 2020JSYJD01) and Shanghai Science and Technology Plan Project (No. 21DZ1204800).

\bibliographystyle{splncs04}
\bibliography{ref}






\end{document}